# The Cherenkov radiation in the galaxy's halo of dark matter

**L M Chechin**

*V.G.Fessenkov Astrophysical Institute,*
*050020, Almaty, Kamenskoye plato*
*National Centre for the Space Researches and Technology of the National Space Agency*
*Republic of Kazakhstan,*

E-mail: chechin-lm@mail.ru; leonid.chechin@gmail.com

**Abstract.** The effect of light refraction in a galaxy's halo of dark matter, described by profiles of Navarro-Frenk-White and Burkett, was considered. Powers of the Cherenkov radiation for the refractive indexes of gravitational fields with these profiles were calculated. It was shown that correspondent radiation temperature in the X-rays range have the magnitude about $T \sim 10(-6) K$. It was also marked that its detection may be the criteria on choosing the preferable dark matter density distribution in a galaxy.

## 1. Introduction

The substance refractive index, as it well known, is the value that equals to ratio between light's phase velocity in vacuum and light's velocity in a given media. However, if any charge particle whose group velocity $v$ in a media is larger than the phase velocity in this substance $c$, i.e. $v > c$ the specific electromagnetic radiation appears. The theory of such radiation – the Cherenkov radiation – was given in the references (Bolotovskij 1957; Ginzburg 1975), for example.

The generalization of Cherenkov radiation on a gravitational field presence was given in monograph (Ivanitzkaya 1979). For the Schwarzschild field this radiation characterizes by upper boundary of frequency $\omega = pc\sqrt{1-\frac{2m}{r}}\frac{r}{m}\left(1-\frac{c}{v}\sqrt{1-\frac{2m}{r}}\right)$. Note that this effect is very small and may be detected in the very strong gravitational field. (It is necessary cite the article (Gupta 1999) where it was shown that if a charged particle moves in the external gravitational field along the geodesic line it will create the Cherenkov radiation also, but for the case of field's nonzero Ricci tensor.)

By the way, the problems of searching the electromagnetic radiation and light rays behavior not only in the local weak gravitational fields, but on the cosmological scales also, where got the essential development at last time. For this confirmation it's enough to mention the searching on the gravitational microlensing by stellar halo of dark matter. In spite of the minuteness of angle images separation (about 0,"001) it is possible to detect the stellar brightness changing due to the mutual movement of star, galaxy and observer at the microlensing process (see, for example, (Paczyński 1996; Zakharov and Sazhin 1996)).

Another aspects of the dark matter searching on the cosmological scales based on the different radiation processes studying in it, including the dynamics of cosmic rays in dark matter, where done by Sizun et al. (2006), Premadi et al. (2001), Moiseev and Proumo (2009), de Boer (2009), Shlomo and Arnon (2009)

Moreover, the searching of dark matter's properties on the small space scales where done in the following articles (Gilmore et al. 2007; Salucci et al. 2010). For example, the dark matter density in the vicinity of Sun was discussed by Salucci et al (2010); the influence of dark matter on the light rays' propagation in the Solar system was considered by Hideyoshi. In fact, in his article (2008) the time delay effect (analog of the Shapiro effect) and the shift of light rays' frequency effect at the spherical-symmetric dark matter distribution where considered. But the shape of dark matter profile was chosen a priori as $\rho(t,r) = \rho(t)\left(\frac{\ell}{r}\right)^k$, where $\rho(t) \approx \rho_0 + \left(\frac{d\rho}{dt}\right)_0 (t-t_0)$. Here $\ell$ and $\rho_0$ are some constant parameters posses by the initial distance and dark matter density physical essences. As for the index of dark matter density's profile $k$ it was specified as -1, 0, and +1.

In spite of the numerical minuteness of these effects (for example, shift frequency is $\Delta\omega \sim 10(-25) Hz$ just at the upper magnitude of dark matter density $\rho_0 \sim 10(-16) g/cm^3$) their possible physical application was focused on the astronomical unit rectification.

In our article the effect of light rays refraction in the dark matter's halo for its some profiles is considered. Beside the powers of Cherenkov radiation for optical, ultraviolet and X-ray frequency ranges are estimated and the possibility of their detecting is discussed also.

## 2. The spherical-symmetric model of dark matter halo

The basic morphological elements of a galaxy are - the nucleus, balge, gaseous-dust disk, clouds of neutral hydrogen, stars cluster and galactic halo. Later on we'll focus our attention on the galactic halo only, because it contains the most part of galaxy's mass and consists of the dark matter preferably (Helfer 2004; Clowe et al. 2006; Strigari et al. 2007).

Let's consider the spherical-symmetric model of dark matter halo. The spherical-symmetric interval of the gravitational field we write down in the standard form (Zel'dovich and Novikov 1971)

$$ds^2 = e^{\nu(r)} \cdot c^2 dt^2 - e^{\lambda(r)} \cdot dr^2 - r^2\left(d\theta^2 + \sin^2\theta \cdot d\varphi^2\right), \qquad (1)$$

where

$$e^{\lambda(r)} = \left(1 - \frac{8\pi G}{c^2 r} \int_0^r \rho(r) r^2 dr \right)^{-1} \quad (2)$$

and

$$e^{\nu(r)} = \exp \int_r^\infty \left[ \frac{8\pi G}{c^4} \left(c^2 \rho(r) + p(r)\right) r \cdot e^{\lambda(r)} - \frac{d\lambda}{dr} \right] dr. \quad (3)$$

In expressions (2) and (3) $\rho(r)$ and $p(r)$ are the mass-energy density and pressure of dark matter, accordingly. The real expressions for mass-energy density, pressure and other characteristics of dark matter distributions are possible get by searching the rotational curves of galaxies. Meanwhile, the plausible space distributions for these physical values are possible get by the N-body simulation method. In fact, the profiles for dark matter energy (Rasia et al. 2004), temperature profiles (Markevitch et al. 1997) and some other characteristics of dark matter were got by this method. In following as the current variable $\rho(r)$ in (1) – (3) we choose two profiles: Navarro-Frenk-White (1996) and Burkett (1995), only. Concerning the pressure we set $p(r) = 0$ that relates to the well-known model of dark matter as an ideal fluid (see, for example, (Böhmer and Harko 2007)).

- Let's write down the Navarro-Frenk-White profile

$$\rho(r) = \frac{\rho_0}{\frac{r}{r_0}\left(1 + \frac{r}{r_0}\right)^2} \quad (4)$$

Assuming that $r_0$ are the sizes of galaxy, we may introduce the small parameter $\frac{r}{r_0} \ll 1$. Putting (4) into (2) and (3) and taking into account terms up to the $\frac{\pi G}{c^2} \rho_0 r_0^2 \left(\frac{r}{r_0}\right)$ order of minuteness, we get

$$e^{\lambda(r)} \approx 1 + \frac{4\pi G}{c^2} \rho_0 r_0^2, \quad (5)$$

$$e^{\nu(r)} \approx 1 + \frac{8\pi G}{c^2} \rho_0 r_0^2 - \frac{4\pi G}{c^2} \rho_0 r_0^2 \left(\frac{r}{r_0}\right). \quad (6)$$

Assuming that in (1) angle $\theta = 0$ and passing in this plane to the Cartesian coordinates ($x^1 = r \cos\varphi$, $x^2 = r \sin\varphi$), with the needed accuracy we get

$$ds^2 = \left(1 + \frac{8\pi G}{c^2}\rho_0 r_0^2 - \frac{4\pi G}{c^2}\rho_0 r_0^2 \left(\frac{r}{r_0}\right)\right) c^2 dt^2 - \left(\delta_{ik} + \frac{4\pi G}{c^2}\rho_0 r_0^2 \frac{x^i x^k}{r^2}\right) dx^i dx^k, \quad (7)$$

where $i, k = 1, 2$. As for the isotropic geodesic line the interval equals zero, it is easy to find the gravitational field's refractive index from (7) -

$$n(r) \approx 1 + \frac{4\pi G}{c^2} \rho_0 r_0^2 \left(1 + \frac{r}{r_0}\right) \quad (8)$$

by introducing the speed of light in a media $v^i = \dfrac{dx^i}{dt}$.

- The Burkett profile in the same designations is

$$\rho(r) = \dfrac{\rho_0}{\left(1 + \dfrac{r}{r_0}\right)\left(1 + \dfrac{r^2}{r_0^2}\right)} . \qquad (9)$$

Later on, as in the previous subsection, we'll take into account terms of $\dfrac{\pi G}{c^2} \rho_0 r_0^2 \left(\dfrac{r}{r_0}\right)$ order, only. Hence, with the needed accuracy we get

$$e^{\lambda(r)} \approx 1, \qquad (10)$$

and

$$e^{\nu(r)} \approx 1 + \dfrac{8\pi G}{c^2} \rho_0 r_0^2 \left(\dfrac{r}{r_0}\right). \qquad (11)$$

According these coefficients the interval of spherical-symmetric gravitational field, produces by the Burkett profile, takes on the form

$$ds^2 = c^2 dt^2 - \left(\delta_{ik} + \dfrac{4\pi G}{c^2} \rho_0 r_0^2 \dfrac{x^i x^k}{r_0 r}\right) dx^i dx^k . \qquad (12)$$

From (12) we get the final expression of refractive index for the corresponding gravitational field

$$n(r) \approx 1 + \dfrac{4\pi G}{c^2} \rho_0 r_0^2 \left(\dfrac{r}{r_0}\right). \qquad (13)$$

**3. The Cherenkov radiation in a galaxy's halo of dark matter**

The power of Cherenkov radiation on a frequency $\omega$ describes by expression (Bolotovskij 1957; Ginzburg 1975)

$$\dfrac{dE}{dt} = W = -\dfrac{q^2 v}{c^2} \int_\omega \left(1 - \dfrac{c^2}{n(\omega)^2 v^2}\right) \omega d\omega , \qquad (14)$$

where $n(\omega)$ is the dispersion refractive index, $q$ is the total charge of particles that moves with mean velocity $v$. As in our cases the refractive index depends on a current coordinate $r$, but not on a frequency $\omega$ (the dispersion absence) than expression (14) simplifies

$$\dfrac{dE}{dt} = W = -\dfrac{q^2 v}{2c^2} \left(1 - \dfrac{c^2}{n(r)^2 v^2}\right) \omega^2 . \qquad (15)$$

That is why the last expression be applied to the refractive indexes (8) and (13) take on the forms

$$\dfrac{dE}{dt} = W_{NFW} \approx -\dfrac{q^2 v}{2c^2} \left(1 - \dfrac{c^2}{v^2} + \dfrac{8\pi G}{v^2} \rho_0 r_0^2 + \dfrac{8\pi G}{v^2} \rho_0 r_0^2 \left(\dfrac{r}{r_0}\right)\right) \omega^2 , \qquad (16)$$

$$\frac{d\mathrm{E}}{dt} = W_B \approx -\frac{q^2 v}{2c^2}\left(1 - \frac{c^2}{v^2} + \frac{8\pi G}{v^2}\rho_0 r_0^2 \left(\frac{r}{r_0}\right)\right)\omega^2 . \tag{17}$$

For the following analyses the gravitational analog of Cherenkov radiation we need to attract the well-known mechanisms of charged particles creation in a galaxy and to estimate the magnitude of probable charge that comes down to the Earth.

For doing this let's consider the possibility of solar emission and radiation from Wolf-Rayet stars usage. It is known that solar wind – the flux of ionized particles (protons, electrons and $\alpha$-particles) - flows from the solar corona. The slow solar wind has the velocity $v \sim 4 \cdot 10(5) m/s$, while the velocity of fast solar wind is $v \sim 7 \cdot 10(5) m/s$.

On the average Sun emits about $1.3 \times 10(36)$ particles per second (Kallenrod 2004; Suess 1999; Carroll 1995) due to the wind. In spite of the vast number of flying particles this type of emission couldn't be used for estimation the Cherenkov radiation, because their velocities are radically smaller than speed of light and their flux is small, also.

However, the energy range of the solar cosmic rays at solar flare is about $10(5) \div 10(11) eV$ that corresponds to their maximal velocities order of $10(6) m/s$. Moreover, in this case is the essentially stronger and the cosmic ray flux - its magnitude may be about $\eta = 10(6) cm^{-2} \sec^{-1}$ of order. But the dark matter does not influent on the cosmic rays propagation practically because the region of their movement is limited by sizes of the solar system.

More active in the aspect of stellar wind emission are the Wolf-Rayet stars. Their wind emits about three Sun masses per one million years. Therefore the rate of substance loss for Wolf-Rayet stars is $\dot{M} \sim 10(20) g/s$ that at seven orders larger than the rate of a substance loss for the stellar wind. Beside, according to (Barlow et al. 1981; Aloisio et al. 2009) the brightness for such type of stars at 10(5) – 10(6) times larger than brightness of Sun. That is why the velocities of ionizing particles from these stars are essentially larger than from the quiet Sun and enrich magnitudes about $10(7) m/s$.

Moreover, a particles in the galactic space, be undergo the additional acceleration by shock waves, acquire the high and ultrahigh energies of order $10(15) \div 10(19) ev$ (Vasiliev and Zelnikov 2008). That is why their final velocities $v \sim 10(7) m/s$ become comparable with the speed of light. And though the number of such particles is highly small - one particle per meter squared during one year on the Earth - they may create the Cherenkov radiation at their traverse through the dark matter in principle.

Examination of expressions (16) and (17) leads to the significant conclusion that from physical viewpoint the most interesting is the first one of them because it includes the largest term $\frac{8\pi G}{v^2}\rho_0 r_0^2$ depending on the central magnitude of the dark matter density $\rho_0$. That is why below we'll focus our searching on this term and estimate the magnitude of it, only.

For doing this we must set the density $\rho_0$ and take into account the following natural restriction for it - $\frac{8\pi G}{v^2}\rho_0 r_0^2 < 1$. Note that in article (Hideyoshi 2008), mentioned above, the upper magnitude of cold dark matter's density is about $\rho \sim 10(-16) g/cm^3$. But such magnitude seems too large because from the mentioned restriction it must be $\rho_0 < 10(-18) g/cm^3$. Remember that mean magnitude of dark matter's density in the Universe is $\rho \sim 0.3\Omega \sim 10(-30) g/cm^3$. Hence, let the following interval for the dark matter's density takes place

$$10(-30) g/cm^3 \leq \rho_{DM} < 10(-18) g/cm^3 . \tag{18}$$

As $\frac{G}{c^2} \sim 10(-28)\,cm/g$, the sizes of typical galaxy are $r_0 \sim 50 \cdot 10(3)\,pc \sim 5 \cdot 10(22)\,cm$ and assuming $\rho_0 \sim 10(-18)\,g/cm^3$ we get the following interval for the above mentioned term

$$10(-12) \leq 8\pi \frac{G}{c^2} \rho_0 r_0^2 < 10(0). \quad (19)$$

From (8) it follows that velocity of charged particle has the expression

$$v \approx c\left(1 - \frac{4\pi G}{c^2} \rho_0 r_0^2\right). \quad (20)$$

Hence the interval of its numerical magnitudes is

$$c(1 - 10(-12)) \leq v \leq c. \quad (21)$$

Basing on the above argumentation we put that charged particles in halo of dark matter can be accelerated up to the velocities $v \sim 10(7)\,m/s$, also. Hence, they will correspond to velocities of ionized particles from the Wolf-Rayet stars and satisfy the relation $\frac{v}{c} \sim 10(-1)$. From these statements we get the magnitude of the central part of dark matter's density

$$\rho_0 = \frac{1 - 10(-1)}{\frac{8\pi G}{c^2} r_0^2} \sim 10(-19)\,g/cm^3. \quad (22)$$

This estimation is at two orders smaller then the corresponding value argued by Hideyoshi (Hideyoshi, 2008). But last value, as it seems, is overvalued because the charged particle's velocity will tent to the speed of light if the density will be $\rho = \rho_0 \sim 10(-16)\,g/cm^3$. According this cause is overvaluing and magnitude of the dark matter's density central part $\rho_0 \sim 10(-14)\,g/cm^3$ that was argued by Vasiliev and Zelnikov (Vasiliev and Zelnikov 2008). Here we are not concerning the ultra light particles existence (Ginzburg 1975) and the results of novel experiment on their detecting (Opera Collaboration 2011).

Now basing on these data we can estimate the possible magnitude of the Cherenkov radiation in the halo of dark matter for the different frequency ranges - visible, ultraviolet and X-rays.

Putting that the flux of high energy charged particles is uniform and isotropic it is possible to estimate the charge quantity that falls down on the Earth during one year. Setting that Earth's radius is $R_E \sim 10(9)\,cm$, area of its hemisphere is $S_E \sim 10(17)\,cm^2$, the rate of particles changing will be $\dot{N} = \dot{\eta} S_E \sim 10(6)\,\sec^{-1}$. Next, a charge particle moves with the velocity $v \sim 10(7)\,m/s$, that is closed to the speed of light, will cross a galaxy during $t = \frac{r_0}{v} \sim 10(13)\,\sec$. Hence, the number of upcoming charge particles will be $N = \dot{N} \cdot t_0 \sim 10(19)$ and the entirety charge corresponding them is about $q = eN \sim 1C$.

As we are interesting the dark matter influence on the Cherenkov radiation, the sum of two first terms in (17) at velocities closed to the speed of light will tend to zero. Hence, the sought-for power radiation is

$$-\frac{dE}{dt} = W_{NFW} \approx 4\pi G \frac{q^2}{c^2 v} \rho_0 r_0 \omega^2. \quad (23)$$

Putting the orders of frequencies that are relate to the visible ($\omega \sim 10(15)\sec^{-1}$), the ultraviolet ($\omega \sim 10(17)\sec^{-1}$) and the X-ray ($\omega \sim 10(19)\sec^{-1}$) ranges into (23) we get the following set of power estimations ( $1\frac{C^2}{cm} \sim 10(1) GeV \sim 10(-3) erg$ ) -

$$-\frac{dE}{dt} = W_{NFW} \sim 10(-1) GeV/\sec,$$
$$-\frac{dE}{dt} = W_{NFW} \sim 10(3) GeV/\sec, \qquad (24)$$
$$-\frac{dE}{dt} = W_{NFW} \sim 10(7) GeV/\sec,$$

accordingly.

The largest power of the Cherenkov radiation, as it seems, is for the X-rays range. That is why the energy order of $E \sim 10(20) GeV$ will emit by particles, that crossing a galaxy, during the time $t = \frac{r_0}{v} \sim 10(13)\sec$. Hence, the corresponding energy density in a halo will be $\varepsilon \approx \frac{E}{r_0^3} \sim 10(-47) GeV/cm^3 \sim 10(-44) erg/cm^3$.

Finally, basing on the Stephan - Boltzmann law $\varepsilon = \sigma T^4$, where $\sigma \sim 10(-15) erg/cm^3 \cdot K^4$ (Gurevitch and Chernin 1978), it is possible estimate the temperature of Cherenkov radiation. Calculations show that its largest order (for X-rays) is about $T \sim 10(-6) K$ and it is closed to the temperature fluctuations of cosmic background radiation $\Delta T \sim 10(-5) K$ (Smoot, 2007). Hence, this temperature lies on the threshold of modern astrophysical measurements.

**4. Conclusion**

What is the physical essence have this type of radiation?

Firstly, its detecting may be the criterion for choosing the preferable dark matter distribution in a galaxy. In fact, now there is number of other, except citing above, types of dark matter profiles. They were searched in many articles, for example (Einasto, 2000; Klypin and Kravtsov, 2000; Treu and Koopmans, 2004; Weber and de Boer, 2009; Kazantzidis et al., 2003; Catena and Ullio, 2009), etc. Though all of them are describing the structure of dark matter distribution in a galaxy – its dependency on redshift, mass, sizes, character of the evolution, etc. – it is desirable to have the general shape of such distribution. In its turn it gives the possibility of understanding the universal properties of the dark matter.

Secondly, in article (Chechin et al., 2012) it was shown that refractive index for the Einasto (2000) dark matter profile is smaller than one. This fact is possible to interpret, probably, as the Cherenkov absorption in the dark matter. Hence, we get the additional perspectives for the dark matter searching, because up to the present time it is considering that dark matter affects only gravitationally, and absorption processes in it was searched for the neutrino (Esteban, 1993).

Concerning the Cherenkov radiation it is necessary to mention and some other articles.

So, in articles (Dremin, 2002; Marklund et al., 2005) it was shown that high energy particles at their traversing the Universe through the cosmic microwave background radiation can, in principle, emit the Cherenkov radiation. But authors were shown that the energy threshold for this radiation is extremely high and its intensity would be too low due to the low density of the "relic photons gas" and very weak interaction of two photons.

In article (Chefranov, 2010) this effect was searched in the relic radiation field, also. But the refractive index was calculated in the framework of quantum electrodynamics. As the result it was found that its difference from the vacuum refractive index ($n = 1$) is about $\delta \sim 10(-42)$ and is drastically smaller than in our case - $\delta_{\min} \sim 10(-13)$.

So, the Cherenkov radiation at the cosmological scales is preferable to search in the galaxy's dark matter halo, namely.

The possibility of dark matter properties experimental searching by the Cherenkov telescopes usage was discussed in the recent article (Sánchez-Conde et al., 2011).

**5. Acknowledgement**

I would like to express gratitude to the Ministry of Education and Sciences, Republic of Kazakhstan for supports this searching in the framework of budget program 055, subprogram 101 "Grant financing of the scientific researchers".